\documentclass{article}
\usepackage{graphicx}
\usepackage{amsmath, latexsym}
\usepackage{amssymb}

\newtheorem{proposition}{Proposition}

\begin{document}

\vspace*{3cm} \thispagestyle{empty}
\vspace{5mm}

\noindent \textbf{\Large Timelike Killing Fields and Relativistic Statistical Mechanics}\\

\noindent \textbf{\normalsize David Klein}\footnote{Department of Mathematics, California State 
University, Northridge, Northridge, CA 91330-8313. Email: david.klein@csun.edu.} \textbf{\normalsize and Peter Collas}\footnote{Department of Physics and Astronomy, 
California State University, Northridge, Northridge, CA 91330-8268. Email: peter.collas@csun.edu.}
\\

\vspace{4mm} \parbox{11cm}{\noindent{\small For spacetimes with timelike Killing fields, we introduce a  ``Fermi-Walker-Killing'' coordinate system and use it to prove a Liouville Theorem for an appropriate volume element of phase space for a statistical mechanical system of particles. We derive an exact relativistic formula for the Helmholtz free energy of an ideal gas and compare it, for a class of spacetimes, to its Newtonian analog, derived both independently and as the Newtonian limit of our formula. We also find the relativistic thermodynamic equation of state. Specific examples are given in Kerr spacetime.}\vspace{5mm}\\


}\\
\vspace{6cm}
\pagebreak

\setlength{\textwidth}{27pc}
\setlength{\textheight}{43pc}
\noindent \textbf{{\normalsize 1. Introduction}}\\

\noindent Curvature in relativistic spacetimes corresponds to tidal forces in Newtonian mechanics, but  curvature effects yield more precise information about physical phenomena.  In particular, general relativity should provide corrections to calculations based on Newtonian physics for the statistical mechanical behavior of a gas subject to a gravitational field. 
Such an investigation was undertaken in \cite{CK}, where an approximate Helmholtz Free Energy function was derived, using statistical mechanical methods, for an ideal gas in orbit around the central mass in Schwarzschild spacetime.  For a partial listing and summary of research in relativistic statistical mechanics, we refer the reader to the introduction in \cite{CK}.\\ 

\noindent In this paper, we consider an ideal gas enclosed in a container with coordinates at rest with respect to an observer whose four-velocity is a timelike Killing vector, and we assume further that the gas particles, along with the container, do not signiÞcantly alter the background metric. It should be noted that the term ``ideal gas'' is somewhat misleading in the context of general relativity. This is because a volume of gas subject to no forces is still affected by the curvature of space-time, and this corresponds to a Newtonian gas subject to gravitational, tidal, and in some instances ``centrifugal forces,'' but otherwise ``ideal.'' We provide details for this association in Sections 6 and 8.\\

\noindent The significance of the frame of reference in relativistic statistical mechanics was discussed in Section 5 of \cite{rovelli}. There, in the context of Minkowski spacetime, it was argued that there is no Lorentz invariant thermal state for an ideal gas system. A thermal state is in equilibrium only in  a preferred Lorentz frame, and therefore breaks Lorentz invariance.  In the Lorentz frame in which the container of the gas is at rest, the Lorentz covariant canonical distribution density for an ideal gas reduces to a product of terms of the form $\exp \beta p_{0}$, where $p_{0}$ is the
negative of the energy of a particle (in that frame) and $\beta$ is inverse temperature.  In that frame, the relativistic statistical mechanical behavior may be easily compared to its non relativistic analog.  \\

\noindent In the context of general relativity our requirement that the spacetime possess a timelike Killing field is physically natural. The four-velocity Killing vector for the container determines a time coordinate for the system and thus an energy component of the four-momentum.  As a consequence, comparisons to corresponding non relativistic statistical mechanical formulas are possible.  Moreover, this choice of four-velocity forces the gas system to evolve in such a way that the geometry of spacetime is unchanging along its worldsurface (since the Lie derivative of the metric vanishes in that direction), and thus it is plausible on physical grounds that the particle system will reach equilibrium. \\

\noindent In order to take full advantage of these features, we introduce, in Section 2, a non rotating, orthonormal ``Fermi-Walker-Killing" coordinate system in which the gas container is at rest, and for which  the Killing vector serves as the time axis. In Section 3, we identify the appropriate volume form on phase space, and then prove Liouville's Theorem in Section 4, improving the local, approximate, version of Liouville's theorem established in \cite{CK}. In Section 5, we derive an exact relativistic formula for the Helmholtz free energy of an ideal gas. In contrast to \cite{CK}, the Boltzmann factor in our partition function is an invariant scalar.  In Section 6 we systematically find Newtonian limits for spacetimes and Killing fields satisfying general conditions. We compare the exact relativistic expressions to their Newtonian analogs, derived both independently and as Newtonian limits of our formula in Section 8.  We also find in Section 7 the relativistic thermodynamic equation of state. Concluding remarks are given in Section 9.\\

\noindent \textbf{{\normalsize 2. Fermi-Walker-Killing Coordinates}}\\

\noindent Throughout we use the sign conventions of Misner, Thorne and Wheeler \cite{MTW}. A vector field $V$ in a spacetime $\mathcal{M}$ is said to be Fermi-Walker transported along a timelike path $\sigma$ if $V$ satisfies the Fermi-Walker equations given by,

\begin{equation}
F_{\vec{u}}(V^{\alpha})\equiv \nabla_{\vec{u}}\;V^{\alpha}+\Omega^{{\alpha}}_{\;\,\beta}V^{\beta}=0\,\label{F1}.
\end{equation}
 
\noindent Here  $\vec{u}$ is the four-velocity along $\sigma$ (i.e., the unit tangent vector), $\Omega^{{\alpha}}_{\;\,\beta}=a^{\alpha}u_{\beta}- 
u^{\alpha} a_{\beta}$, and $a^{\alpha}$ is the four-acceleration.  As usual Greek indices run over $0,1,2,3$ and lower case latin over
$1,2,3$. It is well-known and easily verified that $F_{\vec{u}}(\vec{u})=\vec{0}$, and if vector fields $V$ and $W$ are Fermi-Walker transported along $\sigma$, the scalar product $V^{\beta}W_{\beta}$ is invariant along $\sigma$.  Thus, a tetrad of vectors, Fermi-Walker transported along  $\sigma$ and orthonormal at one point on $\sigma$, is necessarily orthonormal at all points on the path. Moreover, such tetrads may be constructed so that one of the orthonormal vectors is the tangent vector $\vec{u}$. \\

\noindent Let $\sigma (\tau)$ denote the parameterization of $\sigma$ by proper time $\tau$, 
and let $e_{0}(\tau)$, $e_{1}(\tau), e_{2}(\tau), e_{3}(\tau)$ be an orthonormal Fermi-Walker transported tetrad along $\sigma$, with $e_{0}=\vec{u}$. The Fermi-Walker coordinates $x^{0}$, $x^{1}$, $x^{2}$, $x^{3}$ relative to this tetrad on $\sigma$ are given by,

\begin{equation}\label{F2}
\begin{split}
x^{0}\left (\exp_{\sigma(\tau)} (\lambda^{j}e_{j}(\tau)\right)&= \tau \\
x^{K}\left (\exp_{\sigma(\tau)} (\lambda^{j}e_{j}(\tau)\right)&= \lambda^{K}, 
\end{split} 
\end{equation}

\noindent where exponential map, $\exp_{p}(\vec{v})$, denotes the evaluation at affine parameter $1$ of the geodesic starting at the point $p$ in $\mathcal{M}$, with initial derivative $\vec{v}$, and it is assumed that the $\lambda^{j}$ are sufficiently small so that the exponential maps in Eq.\eqref{F2} are defined. From the theory of differential equations, a solution to the geodesic equations depends smoothly on its initial data so it follows from Eq.\eqref{F2} that Fermi-Walker coordinates are smooth. Moreover, it follows from \cite{oniell} that there exists a neighborhood of $\sigma$ on which the map $ X =(x^{0}, x^{1}, x^{2}, x^{3})$ is a diffeomorphism onto an open set in $\mathbb{R}^{4}$ and hence a coordinate chart.\\

\noindent  Assume now that $\vec{K}$ is a timelike Killing vector field in a neighborhood of $\sigma$ and the tangent vector to $\sigma$ is $\vec{K}$, i.e., $e_{0}=\vec{u} = \vec{K}$.  The vector field $\vec{K}$ is the infinitesmal generator of a local one-parameter group $\phi_{s}$ of diffeomorphisms. The function $\phi_{s}$ is the flow with tangent vector $\vec{K}$ at each point, and $\sigma(\tau) = \phi_{\tau}(0,0,0,0)$ (where $(0,0,0,0)$ is the origin in Fermi-Walker coordinates).\\

\noindent Define a diffeomorphism $\bar{X}^{-1}$ from a sufficiently small neighborhood of the origin in $\mathbb{R}^{4}$ to a neighborhood $U$ of $\sigma(0)$ in $\mathcal{M}$ by

\begin{equation} \label{k1}
\bar{X}^{-1}(\bar{x}^{0}, \bar{x}^{1}, \bar{x}^{2}, \bar{x}^{3}) \equiv \phi_{\bar{x}^{0}}(X^{-1}(0, \bar{x}^{1}, \bar{x}^{2}, \bar{x}^{3}))
\end{equation}

\noindent Then $\bar{X} =(\bar{x}^{0}, \bar{x}^{1}, \bar{x}^{2}, \bar{x}^{3})$ is a coordinate system on $U$ which we refer to as Fermi-Walker-Killing coordinates.  The following properties are readily verified:\\

\begin{eqnarray} \label{k2}
\begin{split}
(0, \bar{x}^{1}, \bar{x}^{2}, \bar{x}^{3}) &= (0, x^{1}, x^{2}, x^{3})\\
\frac{\partial}{\partial \bar{x}^{i}}\Bigr|_{\bar{x}^{0} =0}&=\frac{\partial}{\partial x^{i}}\Bigr|_{x^{0} =0}\\
\frac{\partial}{\partial \bar{x}^{0}} &= \vec{K}.
\end{split}
\end{eqnarray}

\noindent Thus, in Fermi-Walker-Killing coordinates, the time coordinate $\bar{x}^{0}$ is the parameter of the flow generated by the Killing field $\vec{K}$ with initial positions of the form $(0, x^{1}, x^{2}, x^{3})$ in Fermi-Walker coordinates.\\

\noindent In the sequel, we will make use of both Fermi-Walker and Fermi-Walker-Killing coordinates.  In the case of Fermi-Walker coordinates we designate momentum form coordinates as $p_{\alpha}$ and in the case of Fermi-Walker-Killing coordinates, the momentum form coordinates will be designated as $\bar{p}_{\alpha}$.  Thus, coordinates of the cotangent bundle of $\mathcal{M}$ will be represented as $\{ x^{\alpha},p_{\beta}\}$ or $\{ \bar{x}^{\alpha},\bar{p}_{\beta}\}$.  The metric components in Fermi-Walker coordinates are designated as $g_{\alpha\beta}$ and in Fermi-Walker-Killing coordinates as $\bar{g}_{\alpha\beta}$. Note that $\bar{g}_{\alpha\beta}$ does not depend on $\bar{x}^{0}$ because of Eqs.\eqref{k2}.\\

\noindent \textbf{{\normalsize 3. Phase Space}}\\

\noindent Consider an observer following a timelike path $\sigma$ with four-velocity $\vec{K}$ which is a Killing vector in $\mathcal{M}$, and assume that there is a gas of non interacting test particles enclosed in a container whose coordinates are at rest with respect to this observer.   We assume further that neither the gas nor the container significantly contribute to the gravitational field, so that the metric on $\mathcal{M}$ is effectively unaltered by their presence.   In a neighborhood of $\sigma$, spacetime points may be charted using Fermi-Walker coordinates or Fermi-Walker-Killing coordinates (described in the previous section), whose origin is $\sigma(0)$.\\

\noindent The Hamiltonian for the system of $N$ ideal gas particles is given by
\begin{equation}
H=\sum_{I=1}^{N}H_{\scriptscriptstyle{I}},\;\;\;\;\mbox{where}\;\;\;\;H_{\scriptscriptstyle{I}}
=\frac{1}{2}\bar{g}_{\scriptscriptstyle{I}}^{\alpha\beta}\bar{p}_{\scriptscriptstyle{I}\scriptstyle{\alpha}}
\bar{p}_{\scriptscriptstyle{I}\scriptstyle{\beta}},
\label{x27}
\end{equation}
and where $H_{\scriptscriptstyle{I}}$ is the Hamiltonian for the $I$th particle. With the Hamiltonian expressed in terms of Fermi-Walker-Killing coordinates, the natural time coordinate, and parameter for the dynamics, of the $N$ particle system is $\bar{x}^{0}$. We consider the dynamics in the following section.  In the present section, we derive the basic 6-form for the relevant phase space.\\

\noindent For simplicity we omit  
the subscript $I$ when it 
is clear that we are referring to a one-particle Hamiltonian.  The state of a single  particle
consists of its four spacetime coordinates together  with its four-momentum coordinates
$\{ \bar{x}^{\alpha},\bar{p}_{\beta}\}$, but the one-particle Hamiltonian satisfies,
\begin{equation}
H_{\scriptscriptstyle{I}}=\frac{1}{2}\bar{g}^{\alpha\beta}\bar{p}_{\alpha}\bar{p}_{\beta}=-\frac{m^{2}c^{2}}{2},
\label{a24}
\end{equation}
where, for timelike geodesics, $m$ is the proper mass of the particle.  Eq.\eqref{a24} may be rewritten as,

\begin{equation}
\bar{p}_{0}=\frac{-\bar{g}^{0i}\bar{p}_{i}+\sqrt{(\bar{g}^{0j}\bar{g}^{0k}-\bar{g}^{00}\bar{g}^{jk})\bar{p}_{j}\bar{p}_{k}-\bar{g}^{00}m^{2}c^{2}}}{\bar{g}^{00}}.
\label{a25}
\end{equation}

\noindent This allows us to reduce the dimension of the one particle state space to seven.  Let $\mathbb{P}$ be the sub bundle of the cotangent bundle of spacetime with each  three dimensional fiber determined by (\ref{a25}). The proof of the following proposition is given in \cite{CK}.\\

\begin{proposition}\label{P1}
The volume 7-form 
$\tilde{\omega}$ on $\mathbb{P}$ given by, 
\begin{equation}
\tilde{\omega}=\frac{1}{\bar{p}^{0}}d\bar{x}^{0}\wedge d\bar{x}^{1}\wedge d\bar{x}^{2}\wedge d\bar{x}^{3}\wedge d\bar{p}_{1}\wedge d\bar{p}_{2}\wedge d\bar{p}_{3},\label{a26}
\end{equation}
is invariant  under all coordinate transformations.
\end{proposition}

\noindent Phase space for a single particle is determined by
the space slice, at fixed time coordinate, orthogonal to the observer's four-velocity (along $\sigma(\tau)$), along with the associated momentum
coordinates, i.e., the cotangent bundle of the space-slice at fixed time coordinate $\bar{x}^{0}$. The appropriate volume form is given by the interior product  
$\textbf{\textit{i}}(m\partial/\partial \tau)\tilde{\omega}$ of the four-momentum vector $m\partial/\partial\tau=\bar{p}^{\alpha}\partial/\partial\bar{x}^{\alpha}$ 
with $\tilde{\omega}$. Then,

\begin{equation}
m\textbf{\textit{i}}(\partial/\partial \tau)\tilde{\omega} = d\bar{x}^{1}\wedge
d\bar{x}^{2}\wedge d\bar{x}^{3}\wedge  d\bar{p}_{1}\wedge d\bar{p}_{2}\wedge d\bar{p}_{3}+(d\bar{x}^{0}\wedge\tilde{\psi}),\label{new27'}
\end{equation}

\noindent where $\tilde{\psi}$ is a five-form.  Since $d\bar{x}^{0}=0$ on vectors on phase space (with fixed time coordinate $\bar{x}^{0}$), the restriction of $m\textbf{\textit{i}}(\partial/\partial \tau)\tilde{\omega}$ to the one-particle (six-dimensional) phase space is,

\begin{equation}
m\textbf{\textit{i}}(\partial/\partial \tau)\tilde{\omega}=d\bar{x}^{1}\wedge d\bar{x}^{2}\wedge d\bar{x}^{3}\wedge  d\bar{p}_{1}\wedge d\bar{p}_{2}\wedge d\bar{p}_{3}.\label{new27}
\end{equation}

\noindent It follows from Proposition \ref{P1}
that the 6-form given by Eq.\eqref{new27} is invariant under coordinate changes of the space 
variables with $\bar{x}^{0}$ fixed.  Physically this means that the
calculations that follow below are independent of the choice of the orthonormal triad $e_{1}(\tau), e_{2}(\tau), e_{3}(\tau)$.\\

\noindent \textbf{{\normalsize 4. A Liouville Theorem}}\\

\noindent The phase space volume form for $N$ particles is the product of $N$ copies of Eq.\eqref{new27}, one copy for each particle. We show that this measure satisfies a Liouville theorem.  The phase space, defined in terms of Fermi-Walker-Killing coordinates, is $6N$ dimensional, but for ease of notation, consider instead an ideal gas consisting of $N$ particles in a one-dimensional box; the
generalization to additional degrees of freedom is straightforward.  Label  the coordinate, momentum, and proper time of the $I$th particle,
$(I=1,\dots,N)$ by $\bar{x}_{\scriptscriptstyle{I}}, \bar{p}_{\scriptscriptstyle{I}},
\tau_{\scriptscriptstyle{I}}$, respectively.  We assume that
$\tau_{\scriptscriptstyle{I}}=\tau_{\scriptscriptstyle{I}}(\bar{x}^{0})$, where $\bar{x}^{0}$ is the time coordinate in the Fermi-Walker-Killing coordinate system.  Our argument requires only that the function
$\tau_{\scriptscriptstyle{I}}(\bar{x}^{0})$ be a smooth function of $\bar{x}^{0}$ (in the interval between collisions).\\

\noindent As noted previously, the Hamiltonian, $H_{\scriptscriptstyle{I}}$, governing the dynamics of a single particle given by Eq. \eqref{a24} is independent of $\bar{x}^{0}$. We may thus define the Hamiltonian vector field, $\vec{v}_{H}$, as follows,

\begin{equation}
\vec{v}_{H}={ \everymath{\displaystyle}
\left(
  \begin{array}{c}
     \dot{\bar{x}}_{1} \\
     \rule{0in}{5ex}
     \dot{\bar{p}}_{1} \\
     \rule{0in}{5ex}
     \vdots\\
     \rule{0in}{5ex}
     \dot{\bar{x}}_{N} \\
     \rule{0in}{5ex}
     \dot{\bar{p}}_{N}
  \end{array}
  \right)=
  \left(
  \begin{array}{c}
     \frac{d\bar{x}_{1}}{d\bar{x}^{0}} \\
     \rule{0in}{5ex}
     \frac{d\bar{p}_{1}}{d\bar{x}^{0}} \\
     \rule{0in}{5ex}
     \vdots\\
     \rule{0in}{5ex}
     \frac{d\bar{x}_{N}}{d\bar{x}^{0}} \\
     \rule{0in}{5ex}
     \frac{d\bar{p}_{N}}{d\bar{x}^{0}}
  \end{array}
  \right)=
\left(
  \begin{array}{c}
     \frac{d\bar{x}_{1}}{d\tau_{1}}\frac{d\tau_{1}}{d\bar{x}^{0}} \\
     \rule{0in}{5ex}
     \frac{d\bar{p}_{1}}{d\tau_{1}}\frac{d\tau_{1}}{d\bar{x}^{0}} \\
     \rule{0in}{5ex}
     \vdots\\
     \rule{0in}{5ex}
     \frac{d\bar{x}_{N}}{d\tau_{N}}\frac{d\tau_{N}}{d\bar{x}^{0}} \\
     \rule{0in}{5ex}
     \frac{d\bar{p}_{N}}{d\tau_{N}}\frac{d\tau_{N}}{d\bar{x}^{0}}
  \end{array}
  \right)}.\label{a28}
\end{equation}

\noindent Liouville's theorem, which states that the Hamiltonian system is measure-preserving, 
is established from the following calculation:
\begin{eqnarray}
\mbox{div}\,\vec{v}_{H}&=&\frac{\partial \dot{\bar{x}}_{1}}{\partial \bar{x}_{1}}+\frac{\partial
\dot{\bar{p}}_{1}}{\partial \bar{p}_{1}}+\cdots+\frac{\partial \dot{\bar{x}}_{N}}{\partial \bar{x}_{N}}+\frac{\partial\dot{\bar{p}}_{N}}{\partial \bar{p}_{N}},\nonumber\\
&=&\left[\frac{\partial}{\partial
\bar{x}_{1}}\left(\frac{d\bar{x}_{1}}{d\tau_{1}}\right)+\frac{\partial}{\partial
\bar{p}_{1}}\left(\frac{d\bar{p}_{1}}{d\tau_{1}}\right)\right]\frac{d\tau_{1}}{d\bar{x}^{0}}+\cdots+
\left[\rule{0in}{3ex}\cdots\;\right]\frac{d\tau_{N}}{d\bar{x}^{0}},\nonumber\\
&=&\left[\frac{\partial^{2}H}{\partial \bar{x}_{1}\partial \bar{p}_{1}}-\frac{\partial^{2}H}{\partial
\bar{p}_{1}\partial \bar{x}_{1}}\right]\frac{d\tau_{1}}{d\bar{x}^{0}}+\cdots+
\left[\rule{0in}{3ex}\cdots\;\right]\frac{d\tau_{N}}{d\bar{x}^{0}}=0.\label{a29}
\end{eqnarray}\\

\noindent It now follows that the $N$-fold product of 6-forms of the form, $d\bar{x}^{1}\wedge
d\bar{x}^{2}\wedge d\bar{x}^{3}\wedge d\bar{p}_{1}\wedge d\bar{p}_{2}\wedge d\bar{p}_{3}$, is invariant under the Hamiltonian flow.\\

\noindent \textbf{{\normalsize 5. The Canonical Free Energy}}\\

\noindent Non relativistic statistical mechanics expresses time averages of observables as phase space averages (in which time is no longer a parameter). In this section, we define and calculate the canonical partition function for a relativistic particle system at rest with respect to the Fermi-Walker-Killing coordinate system, developed in previous two sections. For this purpose, we will identify the phase space of a single particle as the cotangent space of the space-slice at a fixed arbitrary time  $\bar{x}^{0}$, and for convenience we take $\bar{x}^{0}=0$.  This is the collection of simultaneous events at time $\bar{x}^{0}=0$ together with associated momenta.\\

\noindent The four velocity of the observer following the path $\sigma(\tau)$ is $\vec{K}$. The components $K^{\alpha}$ of  $\vec{K}$ in Fermi-Walker coordinates, and $\bar{K}^{\alpha}$ in Fermi-Walker-Killing coordinates, on $\sigma(\tau)$ are $(1, 0, 0, 0)$.  The observer, $\sigma(\tau)$, then measures the energy of a particle with four-momentum $\bar{p}$ to be $-\bar{K}^{\alpha}\bar{p}_{\alpha}= -\bar{p}_{0}$ in Fermi-Walker-Killing coordinates, or $-K^{\alpha} p_{\alpha}= -p_{0}$ in Fermi-Walker coordinates. Since the scalar product is invariant under changes of coordinates, and $\vec{K}$ is a Killing vector field,  $-K^{\alpha} p_{\alpha}= -\bar{K}^{\alpha}\bar{p}_{\alpha}$ is constant along any geodesic tangent to the momentum vector in the container of gas.  Thus, it is natural to define the energy of a particle with mass $m$ and momentum $p_{\alpha}$ to be $-K^{\alpha} p_{\alpha}$, but we modify this definition of energy by the addition of the constant $-mc^{2}$, for the purpose of later comparison of relativistic statistical mechanics and classical statistical mechanics (which traditionally does not take account of the rest mass energy). The energy of a particle is then $-K^{\alpha} p_{\alpha}-mc^{2}$, and the total energy of the particles in the container is the sum of the energies over all $N$ particles.\\

\noindent For what follows, note that by virtue of Eq.\eqref{a24}, the energy $-\bar{K}^{\alpha} \bar{p}_{\alpha}-mc^{2}$ of a particle is uniquely determined by $\vec{K}$ and the space components $p_{i}$ of the momentum form.\\

\noindent Let, \[d\mathbf{\bar{x}}=d\bar{x}^{1}d\bar{x}^{2}d\bar{x}^{3}\,,\;d\mathbf{\bar{p}}=d\bar{p}_{1}d\bar{p}_{2}d\bar{p}_{3},\] and 
\begin{equation}
\label{i1b}
d\mathbf{\bar{x}}^{N}d\mathbf{\bar{p}}^{N}=d\bar{x}^{1}_{1}d\bar{x}^{2}_{1}d\bar{x}^{3}_{1}d\bar{p}_{11}d\bar{p}_{12}d\bar{p}_{13}\ldots
d\bar{x}^{1}_{N}d\bar{x}^{2}_{N}d\bar{x}^{3}_{N}d\bar{p}_{N1}d\bar{p}_{N2}d\bar{p}_{N3}.
\end{equation}

\noindent By Liouville's theorem, established in the previous section, and because particle energies, $-\bar{K}^{\alpha}\bar{p}_{\alpha}$, are constants of the motions along geodesics, it follows that for any fixed $\beta$,

\begin{equation}
\label{in1}
e^{\beta\sum_{I=1}^{N}(\bar{K}^{\alpha} \bar{p}_{\alpha}+mc^{2})_I}d\mathbf{\bar{x}}^{N}d\mathbf{\bar{p}}^{N}
\end{equation}

\noindent is an invariant measure on phase space for the $N$ particle system (the index $I$ in Eq.\eqref{in1} labels the particles in the container, as in Eq.\eqref{x27}). In fact it is not difficult to verify that, 

\begin{equation}\label{Lie}
\mathcal{L}_{\partial/\partial\bar{x}^{0}}\left(e^{\beta(\bar{K}^{\alpha} \bar{p}_{\alpha}+mc^{2})}d\mathbf{\bar{x}}d\mathbf{\bar{p}}\right)=\mathcal{L}_{K}\left(e^{\beta(\bar{K}^{\alpha} \bar{p}_{\alpha}+mc^{2})}d\mathbf{\bar{x}}d\mathbf{\bar{p}}\right)=0,
\end{equation}

\noindent where $\mathcal{L}$ denotes Lie derivative. Analogous to the development of nonrelativistic statistical mechanics, this invariance is a partial justification for the choice of \eqref{in1} as phase space measure. We assume that our particle system is in contact with 
a heat bath, and consequently is in thermal equilibrium. Following the usual 
convention, let $\beta=1/kT$, where $T$ is the temperature of the gas in some volume $V$, and $k$ is Boltzmann's constant.\\

\noindent \textbf{Remark 5.1} An alternative convention for temperature may be used here, consistent with one discussed by Tolman in the context of thermodynamics \cite{tolman} (see also \cite{chernikov}). At any position in the gas container, one may identify an observer with four-velocity $\vec{K}/\|\vec{K}\|$. The energy of a particle with momentum $p$ measured by that observer is  $-K^{\alpha} p_{\alpha}/\|\vec{K}\|$.  Then, defining a position dependent equilibrium temperature $\tilde{T}$ by $\tilde{T}= T/\|\vec{K}\|= T/\sqrt{-\bar{g}_{00}}$ coincides with Tolman's formula and results in the same expressions we find below after normalizing by $\beta mc^{2}$.  \\

\noindent Now, for any fixed time coordinate, $\bar{x}^{0}$, we may define the canonical partition function for the gas in terms of Eq.\eqref{in1}. For convenience of calculation, we choose $\bar{x}^{0}=\tau=0$. It then follows from Eq.\eqref{k2} that the phase space measure at the zero time coordinate is equal to the analogous expression in the (unbarred) Fermi-Walker coordinates.\\

\noindent Thus, we define the canonical partition function for the ideal gas by the following integral expression in Fermi-Walker coordinates:

\begin{equation}
\label{i1a}
Z=\frac{1}{N!(2\pi\hbar)^{3N}}\int_{V^{N}}\int_{\mathbb{R}^{3N}}
e^{\beta\sum_{I=1}^{N}(K^{\alpha} p_{\alpha}+mc^{2})_I}d\mathbf{x}^{N}d\mathbf{p}^{N},
\end{equation}

\noindent with notation defined analogously to (\ref{i1b}).\\

\noindent The volume integrals in Eq.\eqref{i1a} have limits of integration determined by the range of Fermi-Walker coordinates, $x^{1}, x^{2}, x^{3}$, that define the volume of the gas, and the momentum integrals may be calculated explicitly.  To carry out that calculation, consider the single-particle canonical partition function,
\begin{equation}
\label{i1}
z\equiv\frac{1}{(2\pi\hbar)^{3}} \int_{V}\int_{\mathbb{R}^{3}}
e^{\beta K^{\alpha}p_{\alpha}+\beta mc^{2}}\;d\mathbf{x}d\mathbf{p},
\end{equation}

\noindent and define,

\begin{equation}
\label{i1aa}
z_{p}\equiv \int_{\mathbb{R}^{3}}
e^{\beta K^{\alpha}p_{\alpha}}\;d\mathbf{p}.
\end{equation}

\noindent Since $\vec{K}$ and $\vec{p}$ are both timelike,

\begin{equation}
\label{i3}
K^{\alpha}p_{\alpha}=-\|\vec{K}\|\|\vec{p}\,\|\cosh \chi,
\end{equation}

\noindent where $\chi$ is the hyperbolic angle between the two vectors.
In Eq.\eqref{i3}, $\|\vec{p}\,\|=mc$, and we may write, $\|\vec{K}\|=\alpha c$, where $\alpha= \alpha (x^{1}, x^{2}, x^{3})$ is a dimensionless function of the space coordinates.  To evaluate Eq.\eqref{i1aa}, we use the coordinate transformation on $\mathbb{R}^{3}$ given by,

\begin{equation}
\begin{split}\label{i4}
p_{1}&=mc\sinh\chi\cos\theta,\\
p_{2}&=mc\sinh\chi\sin\theta\sin\phi,\\
p_{3}&=mc\sinh\chi\sin\theta\cos\phi,
\end{split}
\end{equation}

\noindent where $0 \leqslant \phi < 2\pi$, $0 \leqslant \theta < \pi$, and $\chi \geqslant 0$. The volume element is

\begin{equation}
\label{i7a}
dp_{1}dp_{2}dp_{3}=(mc)^{3}\sinh^{2}\chi\cosh\chi\sin\theta\, d\chi\, d\theta\, d\phi.
\end{equation}

\noindent Eq.\eqref{i1aa} then becomes,
\begin{equation}
\begin{split} \label{i2}
z_{p}&=(mc)^{3} \int_{0}^{2\pi}\int_{0}^{\pi}\int_{0}^{\infty}
e^{-\beta \|\vec{K}\|\|\vec{p}\,\|\cosh \chi}\,\sinh^{2}\chi\cosh\chi\sin\theta\, d\chi\, d\theta\, d\phi\\
&=4\pi (mc)^{3}\int_{0}^{\infty}e^{-\alpha\beta mc^{2} \cosh \chi}\sinh^{2}\chi\cosh\chi \,d\chi\\
&=4\pi (mc)^{3}\int_{1}^{\infty}e^{-\gamma y} y \sqrt{y^{2}-1}\;dy=\frac{4\pi (mc)^{3}K_{2}(\gamma)}{\gamma},
\end{split}
\end{equation}
where $y=\cosh \chi, \gamma=\gamma(x^{1}, x^{2}, x^{3})=\alpha(x^{1}, x^{2}, x^{3})\beta mc^{2}$, and $K_{2}(\gamma)$ is the modified Bessel function of the second kind.  Substituting the above result for $z_{p}$ back in the expression for the single-particle canonical partition function, Eq.\eqref{i1}, we obtain
\begin{equation}
\label{i10}
z=\frac{4\pi (mc)^{3}}{(2\pi\hbar)^{3}}e^{\beta mc^{2}} \int_{V}\frac{K_{2}(\gamma)}{\gamma}d\mathbf{x}.
\end{equation}

\noindent Thus, Eq.\eqref{i1a} may be rewritten as,

\begin{equation}
\label{partition}
Z=\frac{1}{N!}\left[4\pi \left(\frac{mc}{2\pi\hbar}\right)^{3}e^{\beta mc^{2}}\int_{V}
\frac{K_{2}(\gamma)}{\gamma}d\mathbf{x}\right]^{N}.
\end{equation}

\noindent Following the canonical statistical mechanical prescription, the connection to thermodynamics is given by,
\begin{equation}
F(\beta ,V,N)=-\frac{1}{\beta}\ln Z,
\label{a34}
\end{equation}
where $F(\beta ,V,N)$ is the  Helmholtz free energy of the gas. Note that the free energy given by Eq.\eqref{a34} is a function not merely of the volume of the container, but its actual shape via Eq.\eqref{partition}. This is an unavoidable consequence of the non uniformity of the gravitational field.\\

\noindent In the special case of an observer following a timelike geodesic in Minkowski spacetime, the integrand in Eq.\eqref{partition} is constant, and the free energy may be evaluated explicitly. The result is well-known, see e.g., \cite{pauli}.  In that case the observer has a four-velocity in Fermi coordinates given by $\vec{K}=(1,0,0,0)$. With these components the vector field $\vec{K}$ is a Killing field, and it is clear that $\alpha(\lambda)=\lambda\sqrt{-K^{\alpha}K_{\alpha}}\equiv 1$. Thus, from Eq.\eqref{i10},

\begin{equation}
\label{i10'}
z=e^{\beta mc^{2}}\frac{4\pi (mc)^{3}}{(2\pi\hbar)^{3}}\frac{K_{2}(\beta mc^{2})}{\beta mc^{2}}V,
\end{equation}

\noindent and,

\begin{equation}
F(\beta, V,N) = -NkT\ln \left(\frac{e^{\beta mc^{2}}}{N!}\frac{4\pi (mc)^{3}}{(2\pi\hbar)^{3}}\frac{K_{2}(\beta mc^{2})}{\beta mc^{2}}V\right).
\end{equation}

\noindent consistent with \cite{pauli}. The presence of gravity generalizes the special relativistic result so that $\gamma(x^{1}, x^{2}, x^{3})=\alpha(x^{1}, x^{2}, x^{3})\beta mc^{2}$ becomes a function of spatial coordinates, rather than a constant. \\
 
\noindent \textbf{{\normalsize 6. Newtonian Limit of the Relativistic Free Energy}}\\

\noindent For a timelike path $\sigma$ tangent to a Killing vector $\vec{K}$ in a given spacetime, we have defined via Eqs.\eqref{partition} and \eqref{a34} the relativistic Helmholtz free energy of a container of an ideal gas determined by $\sigma$.  With appropriate assumptions (described below), associated to such a statistical mechanical system is an analogous Newtonian (i.e., nonrelativistic) expression for the Helmholtz free energy of the gas.  This associated Newtonian Helmholtz free energy is the Newtonian limit of Eq.\eqref{a34}, i.e., the limit as $c \rightarrow \infty$, where $c$ is the speed of light.\\

\noindent The asymptotic behavior of $K_{2}(\gamma)$ for large argument is given by (see \cite{E53}),
 
\begin{equation}
\label{i11}
K_{2}(\gamma)\sim\sqrt{\frac{\pi}{2\gamma}}\;e^{-\gamma}.
\end{equation}

\noindent and therefore,

\begin{equation}
\label{i12}
z\sim\left(\frac{m}{2\pi\hbar^{2}\beta}\right)^{\frac{3}{2}} \int_{V}\frac{e^{\beta mc^{2}(1-\alpha)}}{\alpha^{3/2}}\;d\mathbf{x}.
\end{equation}

\noindent In order calculate the Newtonian limit, we let
\begin{equation}
\lambda=\frac{1}{c},
\label{a41}
\end{equation}
and we assume that $\alpha=\alpha(\lambda)$ is a smooth function of $\lambda$ (here and below we suppress the dependence of $\alpha$ on $x^{1}, x^{2}, x^{3}$).  Under fairly general conditions (see the following section), $\alpha(0)=1$ and $\alpha'(0)=0$, where the prime denotes differentiation with respect to $\lambda$.  In that case we may write,

\begin{equation}
\alpha(\lambda)= 1+\frac{1}{2}\alpha''(0)\lambda^{2} + O(\lambda^{3})
\label{i13}
\end{equation}

\noindent Combining Eqs.\eqref{i12} and \eqref{i13}, letting $\lambda \rightarrow 0^{+}$, and denoting the dependence of the one particle partition function on $\lambda$ by $z_{\lambda}$, yields,

\begin{equation}
\begin{split}\label{i14}
\lim_{\lambda \rightarrow 0^{+}}z_{\lambda}&=\left(\frac{m}{2\pi\hbar^{2}\beta}\right)^{\frac{3}{2}}\lim_{\lambda \rightarrow 0^{+}} \int_{V}\frac{e^{-\frac{1}{2}\beta m(\alpha''(0)+O(\lambda))}}{(1+\frac{1}{2}\alpha''(0)\lambda^{2} + O(\lambda^{3}))^{3/2}}\;d\mathbf{x}\\
&= \left(\frac{m}{2\pi\hbar^{2}\beta}\right)^{\frac{3}{2}}\int_{V}e^{-\frac{1}{2}\beta m\alpha''(0)}\;d\mathbf{x} \equiv z_{\text{Newt.}},
\end{split}
\end{equation}

\noindent where $z_{\text{Newt.}}$ is the one-particle Newtonian partition function. We note that $\alpha''(0)$ is a function of $x^{1}, x^{2},x^{3}$ and it follows from Eq.\eqref{i14} that $\frac{1}{2}m\alpha''(0)$ may be identified as the Newtonian, nonrelativistic potential energy $U(x^{1}, x^{2},x^{3})$ of a test particle of mass $m$ with coordinates $(0,x^{1}, x^{2},x^{3})$ due to the gravitational field. This potential energy function $U(x^{1},x^{2},x^{3})$ is normalized (by an additive constant) so that $U(0,0,0) = 0$, i.e., the potential energy on the timelike path $\sigma$ is zero.  This is because $\vec{K}$ is the four velocity on $\sigma$ so $\alpha(\lambda) =\|\vec{K}\|/c$ is identically $1$ as a function of $\lambda$ on  $\sigma$, forcing $\alpha''(0) = 0$ there.  Examples are given in section 7.\\

\noindent Returning to the general case of an $N$ particle system, let, 

\begin{equation}
F_{\lambda}(\beta ,V,N)=-\frac{1}{\beta}\ln Z_{\lambda},
\label{a34'}
\end{equation}

\noindent where the subscript indicates dependence on the parameter $\lambda$ and $Z_{\lambda}$ is given by Eq.\eqref{partition}.  Again, assuming $\alpha(0)=1$ and $\alpha'(0)=0$, it now follows that,\\

\begin{equation}
\lim_{\lambda \rightarrow 0^{+}} F_{\lambda}(V,\beta , N)=F_{\text{Newt.}}(V,\beta , N),
\label{N1}
\end{equation}

\noindent where,

\begin{equation}
F_{\text{Newt.}}(\beta ,V,N)=-\frac{1}{\beta}\ln Z_{\text{Newt.}},
\label{N2}
\end{equation}

\noindent and

\begin{equation}\label{a35'}
Z_{\text{Newt.}}=\left(\frac{m}{2\pi\hbar^{2}\beta}\right)^{\frac{3}{2}N}\frac{1}{N!}\left(\int_{V}e^{-\frac{1}{2}\beta m\alpha''(0)}\;d\mathbf{x}\right)^{N}.
\end{equation}

\noindent The difference between the relativistic and associated Newtonian free energies may be computed directly from Eqs.\eqref{partition} and \eqref{a35'}. Thus,

\begin{equation}
\begin{split}\label{difference}
F(\beta ,&V,N)-F_{\text{Newt.}}(\beta ,V,N) = \frac{1}{\beta}\ln \left(\frac{Z_{\text{Newt.}}}{Z}\right)=NkT\ln \left(\frac{z_{\text{Newt.}}}{z}\right)\\
&=nRT\ln \left(\frac{e^{-\beta mc^{2}}}{\beta mc^{2}}\sqrt{\frac{\pi}{2\beta mc^{2}}}\frac{\int_{V}e^{-\frac{1}{2}\beta m\alpha''(0)}\;d\mathbf{x} }{ \int_{V}\frac{K_{2}(\gamma)}{\gamma}d\mathbf{x}}\right),
\end{split}
\end{equation}

\noindent where $R$ is the universal gas constant ($k$ times Avogadro's number), and $n$ is the number of gram-moles of gas. We note that in the preceding equations, the role of volume $V$ is restricted solely to the identification of (constant) limits of integration for the volume integrals.  In the relativistic context, these limits of integration are determined by proper lengths of the dimensions of the container of gas, and in the Newtonian limit the limits of integration are absolute length measurements. \\

\noindent \textbf{{\normalsize 7. Equations of State}}\\

\noindent In this section we find the equations of state for the relativistic gas particle system, and the equation of state for the corresponding Newtonian system of particles subject to the potential energy function $U(x^{1}, x^{2},x^{3})\equiv m\alpha''(0)/2$.  For notational simplicity and for the purpose of comparison, we use different fonts to distinguish the Pressure $\mathcal{P}$ and volume $\mathcal{V}$ for the Newtonian system from the relativistic counterparts $P$ and $V$.\\  

\noindent Following \cite{Martin-Lof}, the equation of state of the Newtonian particle system, subject to the  gravitational potential 
$U(\mathbf{x}) \equiv U(x^{1}, x^{2},x^{3})\equiv m\alpha''(0)/2$, is given by a local version of Boyle's Law, 

\begin{equation}
\beta \mathcal{P}(\mathbf{x})= \rho_{\text{Newt.}}(\mathbf{x}),
\label{pressure1}
\end{equation}

\noindent where $\mathcal{P}(\mathbf{x})$ is local pressure, and $\rho_{\text{Newt.}}(\mathbf{x}) d\mathbf{x}$ is the probabilistic expectation of the number particles\footnote{More formally the integral of $\rho_{\text{Newt.}}(\mathbf{x})$ over a Borel measurable subset of the region containing the gas particles gives the expected number of particles in that set with respect to the Canonical Gibbs measure.}  at position $\mathbf{x}$ given by,

 \begin{equation}
\rho_{\text{Newt.}} (\mathbf{x})= \frac{N \exp (-\beta U(\mathbf{x}))}{\int_{\mathcal{V}}\exp (-\beta U(\mathbf{x})))d\mathbf{x}}.
\label{pressure2}
\end{equation}

\noindent Integrating both sides of Eq.\eqref{pressure2} over the volume gives 

\begin{equation}\label{pressure5}
\langle \mathcal{P}\rangle \mathcal{V}= NkT,
\end{equation}

\noindent where $\langle \mathcal{P} \rangle = \frac{1}{\mathcal{V}}\int_{\mathcal{V}}\mathcal{P}(\mathbf{x})d\mathbf{x}$ is the averaged pressure.\\  

\noindent An analogous equation of state, 

\begin{equation}
\beta P(\mathbf{x})= \rho(\mathbf{x}),
\label{pressure3}
\end{equation}

\noindent holds in the relativistic case.  However, in this case, the volume element is no longer $d\mathbf{x}$, as in the classical case, but rather $d\Omega\equiv \sqrt{\det q_{ij}}d\mathbf{x}$, where $q_{ij}$ is the metric induced by $g_{\alpha\beta}$ on the space slice at fixed time coordinate $\tau=0$, orthogonal to the observer's four-velocity;  $q_{ij}$ is the metric that measures distances within the container of gas particles. The particle number expectation density function, $\rho(\mathbf{x})$, with respect to the space volume measure $d\Omega$ is then given by,

\begin{equation}
\rho (\mathbf{x})= \frac{N K_{2}(\gamma (\mathbf{x}))}{\gamma(\mathbf{x}) \sqrt{\det q_{ij}} \int_{V}(K_{2}(\gamma (\mathbf{x}))/\gamma (\mathbf{x})) \, d\mathbf{x}},
\label{pressure4}
\end{equation}

\noindent where $\gamma (\mathbf{x})\equiv\gamma=\alpha\beta mc^{2}$, as before. Thus, integrating both sides of Eq.\eqref{pressure3} over the volume $V$ with respect to $d\Omega$ gives,

\begin{equation}\label{pressure6}
\langle P\rangle V= NkT,
\end{equation}

\noindent where $\langle P \rangle= \frac{1}{\mathcal{V}}\int_{\mathcal{V}}P(\mathbf{x}) d\Omega$ is 
the relativistic averaged pressure.\\ 

\noindent From Eqs.\eqref{pressure5} and \eqref{pressure6}, it follows that at temperature $T$, the relativistic volume and averaged  pressure are related to their Newtonian counterparts by,

\begin{equation}\label{pressure7}
\langle P\rangle = \frac{\mathcal{V}}{V} \langle \mathcal{P}\rangle=\frac{\int_{\mathcal{V}}1d\mathbf{x}}{\int_{\mathcal{V}}\sqrt{\det q_{ij}}d\mathbf{x}}\langle \mathcal{P}\rangle,
\end{equation}

\noindent where both integrals are evaluated over the same limits of integration, signified by the subscript $\mathcal{V}$ as described in the paragraph following Eq.\eqref{difference}.\\

\noindent As an illustration of Eq.\eqref{pressure7}, consider a container of gas in Schwarzschild spacetime with fixed space coordinates $r=r_{0}, \theta=\pi/2, \phi=0$ (see Eq.\eqref{schwarz1} below).  Using Eq. (53) of \cite{LN79+}, one may readily calculate $\sqrt{\det q_{ij}}$ to third order, and with more work to higher order, but for the sake of concision, we display the result only to $O(2)$, 

\begin{equation}\label{det}
\sqrt{\det q_{ij}}= 1+\frac{GM\lambda^{2}}{6r_{0}^{3}}\left(2\left(x^{1}\right)^{2}-\left(x^{2}\right)^{2}-\left(x^{3}\right)^{2}\right) +O(3).
\end{equation}

\noindent It is apparent that relativistic gravitational effects are negligible unless $GM$ is of the order of $c^{2}$ or higher.\\

\noindent \textbf{{\normalsize 8. Examples in Kerr spacetime}}\\

\noindent In this section we calculate $\alpha(\lambda)$ and the limiting Newtonian potential energy function, $U(x^{1}, x^{2},x^{3})\equiv m\alpha''(0)/2$, the key ingredients in Eqs.\eqref{partition} and \eqref{a35'}, for the cases of circular geodesic orbits in the equatorial plane of Kerr spacetime and for a zero angular momentum observer, or ZAMO.\\

\noindent The Kerr metric in Boyer-Lindquist coordinates is given by,
\begin{equation}
\begin{split}\label{Kerr1}
ds^{2}=&-\left(1-\frac{2\lambda^{2}GMr}{\rho^{2}}\right)\frac{1}{\lambda^{2}}dt^{2}-\frac{4\lambda^{2}GMar\sin^{2}\theta}{\rho^{2}}dtd\phi+\frac{\Sigma}{\rho^{2}}\sin^{2}\theta d\phi^{2}\\
&+\frac{\rho^{2}}{\Delta}dr^{2}+\rho^{2}d\theta^{2},
\end{split}
\end{equation}

\noindent where,
\begin{equation}
\begin{split}\label{Kerr2}
\rho^{2}&=r^{2}+\lambda^{2}a^{2}\cos^{2}\theta,\\
\Delta&=r^{2}-2GM\lambda^{2}r+\lambda^{2}a^{2},\\
\Sigma&=\left(r^{2}+\lambda^{2}a^{2}\right)^{2}-\lambda^{2}a^{2}\Delta\sin^{2}\theta,\\
\end{split}
\end{equation}

\noindent and where $G$ is the gravitational constant, $M$ is mass, and $a$ is the angular momentum per unit mass, and $-GM\lambda\leqslant a \leqslant GM\lambda$.\\ 

\noindent Below we will need  notation for $\Delta$, and $\Sigma$, evaluated at  the specific coordinates $r=r_{0}$, and $\theta=\pi/2$.  For that purpose, we define,
\begin{equation}
\begin{split}\label{Kerr2b}
\Delta_{0}&=r_{0}^{2}-2GM\lambda^{2}r_{0}+\lambda^{2}a^{2},\\
\Sigma_{0}&=\left(r_{0}^{2}+\lambda^{2}a^{2}\right)^{2}-\lambda^{2}a^{2}\Delta_{0}.\\
\end{split}
\end{equation}

\noindent \textbf{Example 1: Circular Geodesic Orbit} The circular orbit in the equatorial plane (see, e.g., \cite{bardeen}, \cite{BS})  with $d\phi/dt >0$ at radial coordinate $r_{0}$, is given by,

\begin{equation}
\label{schwarz2a}
\sigma(t)=\left(t, r_{0}, \pi/2, \frac{t}{\lambda^{2} a + \sqrt{\frac{r_{0}^{3}}{GM}}}\right),
\end{equation}

\noindent When $a>0$, this orbit is co-rotational and when $a<0$ the orbit is retrograde. All circular orbits in the equatorial plane are stable for $ r_{0} > 9GM\lambda^{2}$, independent of $a$, but the co-rotational circular orbit is stable for smaller values, even inside the ergosphere at $r_{0}= 2GM\lambda^{2}$ when $a$ is sufficiently close to $GM\lambda^{2}$ (cf. \cite{bardeen}).\\

\noindent The Killing field tangent to the path (where it is the four velocity) is given by,

\begin{equation}
\label{T32}
\vec{K}=\left(\frac{\lambda^{2} a\sqrt{GM}+ r_{0}^{(3/2)}}{\mathcal{D}},0,0,\frac{\sqrt{GM}}{\mathcal{D}}\right),
\end{equation}

\noindent where,

\begin{equation}
\label{T33}
\mathcal{D}=\sqrt{2\lambda^{2} a\sqrt{GM}r_{0}^{(3/2)}+r_{0}^{3}-3\lambda^{2}GMr_{0}^{2}}.
\end{equation}

\noindent  Using Eqs.\eqref{Kerr1} and \eqref{T32}, the calculation for $\alpha(\lambda)=\lambda\sqrt{-K^{\alpha}K_{\alpha}}$ in Boyer-Lindquist coordinates, and thence $\alpha''(0)$, is straightforward. The result is,

\begin{equation}
\label{T29}
\frac{1}{2}m\alpha''(0)=-\frac{GMm}{r}-\frac{GMmr^{2}\sin^{2}\theta}{2r_{0}^{3}}+\frac{3GMm}{2r_{0}}.
\end{equation}

\noindent This expression may be interpreted as Newtonian potential energy of a single particle in a container of particles in circular orbit with angular velocity,

\begin{equation}\label{angvel}
\dot{\phi}\equiv\frac{d\phi}{dt}=\sqrt\frac{GM}{r_{0}^{3}},
\end{equation}

\noindent where $r_{0}$ is the radius of the orbit. The first term on the right hand side of Eq.\eqref{T29} is the gravitational potential energy, and the third term is an additive constant that forces $\alpha''(0) =0$ when $r=r_{0}$ and $\theta = \pi/2$, i.e., at the origin of coordinates for the rotating container.  The second term is the centrifugal potential energy, more readily recognized when expressed in cylindrical coordinates.  To that end, let $\tilde{\rho} = r\sin\theta$,  with $\phi$ the azimuthal angle, and let $z$ measure linear distance along the axis of rotation.  The magnitude of angular momentum of a uniformly rotating particle of mass $m$ with cylindrical coordinates $(\tilde{\rho},\phi,z)$ is $\ell=m\tilde{\rho}^{2}\dot{\phi}$, and the centrifugal potential energy is then given by,

\begin{equation}\label{angvel2}
-\frac{\ell^{2}}{2m\tilde{\rho}^{2}}= -\frac{GMm\tilde{\rho}^{2}}{2r_{0}^{2}}=-\frac{GMmr^{2}\sin^{2}\theta}{2r_{0}^{3}}.
\end{equation}

\noindent The minus signs in Eq.\eqref{angvel2} take into account the direction of force, away from the central mass.\\ 

\noindent  In order to compute $\alpha''(0)$ in Fermi coordinates, we select the following tetrad vectors in the tangent space at $\sigma(0)$.

\begin{align}
e_{0}&=\left(\frac{\lambda^{2} a\sqrt{GM}+r_{0}^{(3/2)}}{\mathcal{D}},0,0,\frac{\sqrt{GM}}{\mathcal{D}}\right),\notag\\
e_{1}&=\left (0,\frac{\sqrt{\Delta_{0}}}{r_{0}},0,0\right ),\notag\\
e_{2}&=\left (0,0,\frac{1}{r_{0}},0\right )\label{tt1},\\
e_{3}&=\left (\!\frac{\lambda^{2}\sqrt{GM}\!\left(\lambda^{2}a^{2}\!+\!r_{0}^{2}\!-\!2\lambda^{2} a\sqrt{GMr_{0}}\right)}{\mathcal{D}\sqrt{\Delta_{0}}},0,0,\!\frac{\lambda^{2}\sqrt{GM}(a -2\sqrt{r_{0}})+r_{0}^{(3/2)}}{\mathcal{D}\sqrt{\Delta_{0}}}\!\right)\notag
\end{align}

\noindent This tetrad may be extended via parallel transport to the entire circular orbit given by Eq.\eqref{schwarz2}, but we need these tetrad vectors only at $\sigma(0)$. We note that in the case that $a=0$, this is the same tetrad utilized in \cite{CK} for the estimation of the free energy of a gas in circular orbit in Schwarzshild spacetime.\\ 

\noindent Fermi coordinates relative to these coordinate axes (i.e., the above tetrad) may be calculated using Eq.(27) of \cite{KC}. The result for the Boyer-Lindquist coordinates $r$ and $\theta$ to second order expressed in the Fermi space coordinates $x^{1},x^{2},x^{3}$ at ($x^{0} = \tau = t=0$) is,
\begin{eqnarray}
\theta&=&\frac{\pi}{2}+\frac{x^{2}}{r_{0}}-\frac{\sqrt{\Delta_{0}}\,x^{1}x^{2}}{r_{0}^{3}}+ \cdots,\label{tr3}\\
r&=&r_{0}+\frac{\sqrt{\Delta_{0}}\,x^{1}}{r_{0}}+\frac{\left(r_{0}\lambda^{2}GM-\lambda^{2}a^{2}\right)\left(x^{1}\right)^{2}}{2r_{0}^{3}}\nonumber\\
&\quad&+\frac{\Delta_{0}\left(x^{2}\right)^{2}}{2r_{0}^{3}}+\frac{\left(r_{0}-\lambda^{2}GM\right)\left(x^{3}\right)^{2}}{2r_{0}^{2}}+\cdots\label{tr3'},
\end{eqnarray}

\noindent Now, calculating $\alpha(\lambda)=\lambda\sqrt{-K^{\alpha}K_{\alpha}}$ in Boyer-Lindquist coordinates, substituting for $r$ and $\theta$ using Eqs.\eqref{tr3} and \eqref{tr3'} gives,

\begin{eqnarray}
\begin{split}\label{alpha}
\alpha(\lambda) =1&+\frac{GM\lambda^{2}\left(r_{0}^{2}+3\,a^{2}\lambda^{2}-4\,a\lambda^{2}\sqrt{r_{0}\,GM}\right)\left(x^{2}\right)^{2}}{2\,r_{0}^{2}\,\mathcal{D}^{2}}\\
&-\frac{3\,GM\lambda^{2}\Delta_{0}\left(x^{1}\right)^{2}}{2\,r_{0}^{2}\,\mathcal{D}^{2}}+O(3)
\end{split}
\end{eqnarray}

\noindent Computing the second derivative with respect to $\lambda$ at $\lambda=0$ yields,

\begin{equation}
\label{T30}
\begin{split}
\frac{1}{2}&m\alpha''(0)=\left(-\frac{GMm}{r_{0}}+\frac{GMm\,x^{1}}{r_{0}^{2}}-\frac{GMm\left(2\left(x^{1}\right)^{2}-\left(x^{2}\right)^{2}-\left(x^{3}\right)^{2}\right)}{2r_{0}^{3}}\right)\\
&-\left(\frac{GMm}{2r_{0}}+\frac{GMm\,x^{1}}{r_{0}^{2}}+\frac{GMm\left(\left(x^{1}\right)^{2}+\left(x^{3}\right)^{2}\right)}{2r_{0}^{3}}\right)+\frac{3GMm}{2r_{0}}+O(3).
\end{split}
\end{equation}

\noindent Eq.\eqref{T30} may be compared term-by-term with Eq.\eqref{T29}. The expression in the first pair of parentheses on the right hand side of Eq.\eqref{T30} is the Taylor exapansion to second order of the gravitational potential, which is the first term on the right hand side of Eq.\eqref{T29}.  The second terms in both equations are related analogously.  At the point $\sigma(0)$ in the orbit, the Cartesian variable $x^{1}$ in Eq.\eqref{T30} measures (Newtonian) distance from the origin of coordinates in the radial direction away from the central mass, $x^{2}$ measures distance in the ``z direction'' parallel to the axis of rotation, and $x^{3}$ measures distance in the tangential direction, parallel to the motion of the container of gas in orbit. (However, these orientations do not hold at other parts of the orbit since the coordinate axes are nonrotating.)  Eq.\eqref{T30} may obviously be simplified to yield the potential energy function,

\begin{equation}
\label{T31}
U(x^{1},x^{2},x^{3})=\frac{1}{2}m\alpha''(0)=-\frac{3GmM\left(x^{1}\right)^{2}}{2r_{0}^{3}}+\frac{GmM\left(x^{2}\right)^{2}}{2r_{0}^{3}}+O(3).
\end{equation}

\noindent Combining Eq.\eqref{T31} with Eqs.\eqref{N2} and \eqref{a35'} immediately yields the Newtonian free energy for a dilute gas in circular orbit around a central mass to which the relativistic counterpart Eq.\eqref{a34'} (with $\lambda =1/c$) may be compared.\\  

\noindent In the special case that $a=0$, the Kerr metric (Eq.\eqref{Kerr1}) reduces to the Schwarzschild metric,

\begin{equation}
\label{schwarz1}
ds^{2}=-\left(1-\frac{2GM\lambda^{2}}{r}\right)\frac{1}{\lambda^{2}}dt^{2}+\displaystyle\frac{dr^{2}}{\displaystyle
\left(1-\frac{2GM\lambda^{2}}{r}\right)}+r^{2}(d\theta^{2}+\sin^{2}\theta d\phi^{2}).
\end{equation}

\noindent  In this case, Eq.\eqref{T31} may be derived in an alternative manner as follows. The Killing field $\vec{K}$ tangential to the circular geodesic orbit with radial coordinate $r_{0}$ was calculated in \cite{KC} to second order in the Fermi coordinates, using the tetrad given by Eqs.\eqref{tt1} with $a=0$.  With dependence on $G$ and $\lambda$ made explicit, it was found (see Eq.(61) of \cite{KC}) that,

\begin{equation}
\label{schwarz2}
\vec{K}(\tau,x^{1},x^{2},x^{3})=\left(1,\,-\sqrt{\frac{GM}{r_{0}^{3}}}\;x^{3},\,0,\,\sqrt{\frac{GM}{r_{0}^{3}}}\; x^{1}\right)+O(3).
\end{equation} 

\noindent Using Eq.\eqref{schwarz2}, the Newtonian limit of the energy per particle,

\begin{equation}\label{schwarz3}
E=-K^{\alpha}p_{\alpha}-mc^{2} = -\left(p_{0}+\frac{m}{\lambda^{2}}\right) +\sqrt{\frac{GM}{r_{0}^{3}}}\left(x^{3}p_{1}-x^{1}p_{3}\right),
\end{equation}

\noindent may be computed directly. The limit of the first term on the right side of Eq.\eqref{schwarz3} was calculated in \cite{CK} as,

\begin{equation}
\lim_{\lambda\rightarrow 0^{+}}\left(p_{0}+\frac{m}{\lambda^{2}}\right)=-\left(\frac{\textbf{\textit{p}}^{2}}{2m}-
\frac{GMm\left(2(x^{1})^{2}-(x^{2})^{2}-(x^{3})^{2}\right)}{2r_{0}^{3}}\right),
\label{schwarz4}
\end{equation}
where $\textbf{\textit{p}}^{2}=p_{1}^{2}+p_{2}^{2}+p_{3}^{2}$ (and because of the use of different unit conventions, ``$p_{0}$'' in Eq.\eqref{schwarz4} corresponds to ``$p_{0}/\lambda$'' in \cite{CK}). Thus,

\begin{equation}\label{schwarz5}
\lim_{\lambda\rightarrow 0^{+}} E =\frac{\textbf{\textit{p}}^{2}}{2m}-
\frac{GMm\left(2(x^{1})^{2}-(x^{2})^{2}-(x^{3})^{2}\right)}{2r_{0}^{3}}
+\sqrt{\frac{GM}{r_{0}^{3}}}\left(x^{3}p_{1}-x^{1}p_{3}\right).
\end{equation}

\noindent Completing the square for the sum of the first and third terms on the right hand side of Eq.\eqref{schwarz5} gives in the case of $p_{1}$,

\begin{equation}
\begin{split}\label{schwarz6}
\frac{p_{1}^{2}}{2m}+\sqrt{\frac{GM}{r_{0}^{3}}}x^{3}p_{1}&= \frac{1}{2m}\left(p_{1}+ \sqrt{\frac{GM}{r_{0}^{3}}}mx^{3}\right)^{2}-\frac{GmM\left(x^{3}\right)^{2}}{2r_{0}^{3}}\\
&\equiv \frac{\tilde{p}_{1}^{2}}{2m}-\frac{GmM\left(x^{3}\right)^{2}}{2r_{0}^{3}},
\end{split}
\end{equation}

\noindent with a similar expression for the quadratic polynomial in $p_{3}$.  The result is,

\begin{equation}\label{schwarz7}
\lim_{\lambda\rightarrow 0^{+}} E =\frac{\tilde{p}_{1}^{2}}{2m}+\frac{p_{2}^{2}}{2m}+\frac{\tilde{p}_{3}^{2}}{2m}-\frac{3GmM\left(x^{1}\right)^{2}}{2r_{0}^{3}}+\frac{GmM\left(x^{2}\right)^{2}}{2r_{0}^{3}}+O(3), 
\end{equation}

\noindent where,

\begin{equation}\label{schwarz8}
\tilde{p}_{3}= {p}_{3}- \sqrt{\frac{GM}{r_{0}^{3}}}mx^{1}.
\end{equation}

\noindent  Since Lebesgue measure $dp_{i}$ is invariant under translations on $\mathbb{R}$, Eq.\eqref{schwarz7} is consistent with Eq.\eqref{T31}.\\

\noindent \textbf{Example 2: Zero Angular Momentum Observer} The Killing field for a Zero Angular Momentum Observer, or ZAMO, in Boyer-Lindquist coordinates in Kerr spacetime is given by,

\begin{equation}
\label{Z1}
\vec{K}=\left(\frac{1}{r_{0}}\sqrt{\frac{\Sigma_{0}}{\Delta_{0}}},0,0,\frac{2GMa\lambda^{2}}{\sqrt{\Delta_{0}\Sigma_{0}}}\right),
\end{equation}

\noindent Using Eqs.\eqref{Kerr1} and \eqref{T32}, the calculation for $\alpha(\lambda)=\lambda\sqrt{-K^{\alpha}K_{\alpha}}$ yields,

\begin{equation}
\label{Z2}
\alpha(\lambda)=\left[\frac{\left(\rho^{2}-2GM\lambda^{2}r\right)\Sigma_{0}^{2}-4G^{2}M^{2}a^{2}\lambda^{6}r_{0}(r_{0}\Sigma-2r\Sigma_{0})\sin^{2}\theta}{r_{0}^{2}\rho^{2}\Delta_{0}\Sigma_{0}}\right]^{\frac{1}{2}}.
\end{equation}

\noindent The limiting Newtonian potential is readily calculated as,

\begin{equation}
\label{Z3}
\frac{1}{2}m\alpha^{\prime\prime}(0)=\frac{GmM}{r_{0}}-\frac{GmM}{r},
\end{equation}

\noindent as expected. To find the corresponding expressions in Fermi-Walker coordinates, we use the following tetrad of vectors in the tangent space at $\sigma(0)$,

\begin{eqnarray}
e_{0}&=&\left(\frac{1}{r_{0}}\sqrt{\frac{\Sigma_{0}}{\Delta_{0}}},0,0,\frac{2GMa\lambda^{2}}{\sqrt{\Delta_{0}\Sigma_{0}}}\right),\label{Z4}\\
e_{1}&=&\left(0,\frac{\sqrt{\Delta_{0}}}{r_{0}},0,0\right),\label{Z5}\\
e_{2}&=&\left(0,0,\frac{1}{r_{0}},0\right),\label{Z6}\\
e_{3}&=&\left(0,0,0,\frac{r_{0}}{\sqrt{\Sigma_{0}}}\right).\label{Z7}
\end{eqnarray}

\noindent Again using Eq.(27) of \cite{KC} we find the transformation from Boyer-Lindquist to Fermi-Walker coordinates is given by Eq.\eqref{tr3} and,

\begin{equation}
\begin{split}\label{Z8}
r&=r_{0}+\frac{\sqrt{\Delta_{0}}}{r_{0}}x^{1}+\frac{\left(r_{0}GM-a^{2}\right)\lambda^{2}}{2r_{0}^{3}}\left(x^{1}\right)^{2}\\
&\quad+\frac{\Delta_{0}}{2r_{0}^{3}}\left(x^{2}\right)^{2}+\frac{\Delta_{0}\left(r_{0}^{3}-GMa^{2}\lambda^{4}\right)}{2r_{0}^{2}\Sigma_{0}}\left(x^{3}\right)^{2}+O(3).
\end{split}
\end{equation}

\noindent Now, combining Eqs.\eqref{tr3} ,\eqref{Z2}, and \eqref{Z8} gives,

\begin{equation}
\begin{split}\label{Z9}
\alpha (x^{1},x^{2},x^{3},\lambda)&=1+\frac{GM\lambda^{2}\,A}{r_{0}^{2}\sqrt{\Delta_{0}}\;D}\,x^{1}-\frac{GM\lambda^{2}\,B_{1}}{2r_{0}^{4}\;D^{2}}\left(x^{1}\right)^{2}\\
&\quad+\frac{GM\lambda^{2}\,B_{2}}{2r_{0}^{4}\;D}\left(x^{2}\right)^{2}+\frac{GM\lambda^{2}\,B_{3}}{2r_{0}^{4}\;D^{2}}\left(x^{3}\right)^{2}+O(3),
\end{split}
\end{equation}

\noindent where
\begin{equation}
\begin{split}\label{Z10}
&A=r_{0}^{4}+2r_{0}^{2}a^{2}\lambda^{2}+a^{4}\lambda^{4}-4GMr_{0}a^{2}\lambda^{4},\\
&B_{1}=2r_{0}^{7}+7r_{0}^{5}a^{2}\lambda^{2}+8r_{0}^{3}a^{4}\lambda^{4}+11GMr_{0}^{4}a^{2}\lambda^{4}+3r_{0}a^{6}\lambda^{6}\\&\quad+14GMr_{0}^{2}a^{4}\lambda^{6}+7GMa^{6}\lambda^{8}-4G^{2}M^{2}r_{0}a^{4}\lambda^{8},\\
&B_{2}=r_{0}^{4}+4r_{0}^{2}a^{2}\lambda^{2}+3a^{4}\lambda^{4}-4GMr_{0}a^{2}\lambda^{4},\\&B_{3}=\left(r_{0}^{3}-GMa^{2}\lambda^{4}\right)A,\\
&D=r_{0}^{3}+r_{0}a^{2}\lambda^{2}+2GMa^{2}\lambda^{4}.
\end{split}
\end{equation}

\noindent Calculating $\alpha^{\prime\prime}(\lambda=0)$ gives the limiting Newtonian potential energy function,

\begin{equation}
\label{Z11}
\frac{1}{2}m\alpha^{\prime\prime}(0)=\frac{GmM}{r_{0}^{2}}x^{1}-\frac{GmM}{r_{0}^{3}}\left(x^{1}\right)^{2}+\frac{GmM}{2r_{0}^{3}}\left(x^{2}\right)^{2}+\frac{GmM}{2r_{0}^{3}}\left(x^{3}\right)^{2}+O(3),
\end{equation}

\noindent which is the Taylor expansion of the Newtonian potential energy function $-GMm/r$ plus an additive constant.\\ 

\noindent \textbf{{\normalsize 9. Conclusions}}\\

\noindent For spacetimes with timelike Killing fields, we have developed the canonical ensemble for an ideal gas whose container is at rest relative to what we call ``Fermi-Walker-Killing'' coordinates.  The Helmholtz free energy for such a system is given by Eqs.\eqref{partition} and  \eqref{a34}.\\ 

\noindent The Boltzmann factor in these expressions is invariant with respect to coordinate changes, so one may use an arbitrary coordinate system, with the appropriate transformation of the volume form for phase space in the integral expressions of our formulas.   However, the Fermi-Walker-Killing coordinate system is the natural choice for statistical mechanics because the container for the particle system is at rest in this frame, and these coordinates lead to a general formula for Newtonian limits, for the purpose of comparison. Under fairly general conditions (namely, $\alpha(0)=1$ and $\alpha'(0)=0$), the Newtonian limits in these spacetimes are given by Eqs.\eqref{N2} and \eqref{a35'}. Because of the inhomogeneity of the geometry of space in the relativistic case, and the corresponding non translation invariance of the gravitational potential energy function in the Newtonian analogs, pressure is necessarily a function of position. The relationship between the relativistic and Newtonian pressures, averaged over volumes, is given by Eq.\eqref{pressure7}.\\

\noindent The generalization of our results to a gas with interactions might be carried out by starting with a covariant Lagrangian that incorporates the interaction between the particles in the container.  An energy momentum tensor determined by such a Lagrangian together with the Einstein field equations then in principle determines the metric.  From there the metric together with a timelike Killing field (or perhaps an ``approximate Killing field''), following the formalism of this paper, may be used to develop the canonical ensemble.  A simpler, approximate theory for nonrelativistic particles with interactions, for the purpose of studying the effects of the background curvature could be achieved, following the classical formulations, by including a potential energy function of the proper distance between particles in the Boltzmann factor in the partition function.\\

\noindent However, an important physical observation is already available from the results of this paper for noninteracting particles.   As noted earlier, relativistic corrections to the pressure given by Eq.\eqref{pressure7} become significant (at arbitrary temperature) only in the presence of extremely large masses.  It is well-known that special relativistic corrections to the Helmholtz free energy for the ideal gas are insignificant except at extremely high temperature \cite{pauli}.  It follows from our results for the general relativistic case, in particular Eq.\eqref{difference}, that curvature corrections from strong gravitational fields (so that $\alpha$ varies rapidly as a function of position) to the free energy become significant only at temperatures of the same magnitude as in the special relativistic case.\\

\end{document}